\documentclass[aps,prd,preprint,superscriptaddress,showpacs]{revtex4}
\usepackage{graphicx}

\usepackage{ulem}
\usepackage{color}
\definecolor{My_red}        {cmyk}{0.00,1.00,1.00,0.20}

\newcommand{\bmat}{\left(\begin{array}}
\newcommand{\emat}{\end{array}\right)}
\newcommand{\beq}{\begin{equation}}
\newcommand{\eeq}{\end{equation}}
\newcommand{\wt}{\widetilde}

\newcommand{\K}{\mbox{$\not \hspace{-0.10cm} Z_2$ }}
\newcommand\ignore[1]\relax

\def\ra{\rightarrow}

\def\Ld{\Lambda}
\def\ld{\lambda}
\def\f{\frac}

\def\bwt{\begin{widetext}}
\def\ewt{\end{widetext}}
\def\be{\begin{equation}}
\def\ee{\end{equation}}
\def\bea{\begin{eqnarray}}
\def\eea{\end{eqnarray}}
\def\bean{\begin{eqnarray*}}
\def\eean{\end{eqnarray*}}
\def\bary{\begin{array}}
\def\eary{\end{array}}
\def\bit{\begin{itemize}}
\def\eit{\end{itemize}}

\def\ra{\rightarrow}

\def\Ld{\Lambda}

\def\ld{\lambda}

\def\su5u1{SU(5) \times U(1)}
\def\fsu5u1{SU(5) \times U(1)'}
\def\so10{SO(10)}
\def\sq20{SO(10) \times SO(10)}

\def\ra{\rightarrow}

\def\Ld{\Lambda}
\def\ld{\lambda}
\def\f{\frac}

\def\L{\left(}
\def\R{\right)}

\def\ra{\rightarrow}

\def\Ld{\Lambda}

\def\ld{\lambda}

\def\su5u1{SU(5) \times U(1)}
\def\fsu5u1{SU(5) \times U(1)'}
\def\so10{SO(10)}
\def\sq20{SO(10) \times SO(10)}

\usepackage[centertags]{amsmath}
\usepackage{amssymb}

\begin{document}

\title{Asymmetric Origin for Gravitino Relic Density
in the Hybrid Gravity-Gauge Mediated Supersymmetry Breaking}

\author{Zhaofeng Kang}

\email{zhfkang@itp.ac.cn}

\affiliation{ State Key Laboratory of Theoretical Physics,
      Institute of Theoretical Physics, Chinese Academy of Sciences,
Beijing 100190, P. R. China }

\author{Tianjun Li}

\email{tli@itp.ac.cn}

\affiliation{ State Key Laboratory of Theoretical Physics,
      Institute of Theoretical Physics, Chinese Academy of Sciences,
Beijing 100190, P. R. China }

\affiliation{George P. and Cynthia W. Mitchell Institute for
Fundamental Physics, Texas A$\&$M University, College Station, TX
77843, USA }

\date{\today}

\begin{abstract}

We propose the hybrid gravity-gauge mediated supersymmetry breaking where
the gravitino mass is about several GeV. 
The strong constraints on supersymmetry viable parameter space
from the CMS and ATLAS experiments at the LHC can be relaxed due to the
heavy colored supersymmetric particles, and it is consistent with null
results in the dark matter (DM) direct search experiments such as
XENON100. In particular, the possible maximal flavor and CP violations
from the relatively small gravity mediation may naturally account for
the recent LHCb anomaly. In addition, because the gravitino mass is
around the asymmetric DM mass, we propose the asymmetric origin of
the gravitino relic density and solve the cosmological coincident
problem on the DM and baryon densities $\Omega_{\rm DM}:\Omega_{B}\approx 5:1$.
The gravitino relic density arises from asymmetric metastable particle
(AMP) late decay. However, we show that there is no AMP candidate in the
minimal supersymmetric Standard Model (SM) due to the robust gaugino/Higgsino
mediated wash-out effects. Interestingly, AMP can be realized in the well
motivated supersymmetric SMs with vector-like particles or continuous
$U(1)_R$ symmetry. Especially, the lightest CP-even Higgs boson mass
can be lifted in the supersymmetric SMs with vector-like particles.

\end{abstract}
\pacs{12.60.Jv, 14.70.Pw, 95.35.+d}

\maketitle

\section{Introduction and Motivation}

The most natural solution to the gauge hiearchy problem is
supersymmetry (SUSY). In the Minimal Supersymmetric Standard Model (MSSM)
with $R$-parity,  the Standard Model (SM) gauge couplings can
be unified at about $2\times 10^{16}$~GeV, which strongly indicates
the Grand Unified Theories (GUTs). And there is a dark matter (DM) candidate,
which is the lightest supersymmetric particle (LSP) such as neutralino,
sneutrino, or gravitino, etc. Thus, the supersymmetric SMs (SSMs) are
anticipated among the most promising new physics beyond the SM.
However, the CMS~\cite{arXiv:1109.2352} and
ATLAS~\cite{arXiv:1110.2299} experiments at the LHC have given strong
constraints on viable supersymmetry parameter space, in particular,
the colored supersymmetric particles (sparticles) such as squarks and
gluinos must be heavy at least
around the 1 TeV or larger. In addition, the LHCb experiment~\cite{LHCb} have
recently observed large
$\Delta A_{CP} (A_{CP}(D^0\to K^+ K^-)-A_{CP}(D^0\to \pi^+ \pi^-)$.
Also, afte  years effort the direct DM detection brings confusing results,
as well a strong exclusion line from XEON100~\cite{XENON} that pushes
the lightest neutralino LSP as a DM candidate to
a quite embarrassing circumstances~\cite{Farina:2011bh}.

To understand these results, we may have to re-examine the basic assumptions
underline the experimental searches: (1) the CMS and ATLAS collaborations
have mainly studied the viable supersymmetry parameter space in the mSUGRA
where the squarks and gluinos might be relatively light; (2)
the assumption for the DM direct detection experiments is
 the weakly interactive  massive particle (WIMP) scenario where the
DM density can be reproduced naturally. In fact,
the relatively heavy squarks and gluinos can be realized elegantly in the gauge
mediated supersymmetry breaking (GMSB) (see the review~\cite{GMSBearly}
and references therein), where the gravitino is a dark matter candidate.
Thus, it is not surprised that we have null results at the LHC and
XENON100. However, gauge mediation is flavour blind, and then we can not
explain the LHCb recent results. The simple solution is that we turn on
the gravity mediation which can induce the flavour and CP violations.
Because of the strong constraints on flavour changing neutral current (FCNC)
and on the electric dipole moments for neutron and electron,  such gravity
mediation must be small. Thus, we propose the hybrid gravity-gauge mediated
supersymmetry breaking where the gravitino mass is a few GeV.

Moreover, there is a well-known coincident puzzle between the DM
and baryon densities. In our Universe today, the ratio of the DM and baryonic
matter energy densities is at the order one, {\it i.e.},
 $\Omega_{\rm DM}h^2:\Omega_{B}h^2\approx 5:1$. The asymmetric DM (ADM)
framework~\cite{Barr:1990ca} provide an elegant solution, and it
predicts the dark matter mass around 5 GeV. Thus, if the
gravitino relic density has asymmetric origin, we can solve the
coincident problem. In this paper, we propose that the
gravitino acquires the relic density from asymmetric metastable particle (AMP) late decay.
For simplicity, we called such gravitino as the ``asymmetric" gravitino.
  This  proposal combines   $\wt G$ late decay production~\cite{Feng:2003xh, Feng:2004mt} with the
asymmetric DM (ADM) framework~\cite{Barr:1990ca}.
   And then it inherits both merits.
Especially, just as the ADM the ``asymmetric" gravitino is predicted to
 have a mass around 5 GeV. Therefore, the SUSY breaking should be mediated dominantly
by GMSB but nearly hit the maximal flavor and CP violations from democracy gravity mediation.
In this paper, we will show that there is no AMP candidate in the
MSSM due to the robust gaugino/Higgsino
mediated wash-out effects. Interestingly, AMP can be realized in the well
motivated supersymmetric SMs with vector-like particles or continuous
$U(1)_R$ symmetry. Note that the ATLAS  and CMS Collaborations have reported an excess of events
for the SM-like Higgs boson with mass around $126$~GeV and $124$~GeV,
respectively~\cite{Collaboration:2012si, Chatrchyan:2012tx},
the supersymmetric SMs with vector-like particles are very interesting
since we can lift the lightest CP-even Higgs boson
mass~\cite{Babu:2008ge, Evans:2011uq, Martin:2009bg, Graham:2009gy}.

The paper is organized as follows. In Section II we study the general conditions
that such a scenario  can be realized. In Section III we consider two
typical AMP candidates in the supersymmetric SMs with vector-like particles or continuous
$U(1)_R$ symmetry. Our conclusion is in Section IV.

\section{The Gravitino Relic Density from the AMP Late Decay}

\subsection{The Natural DM: ADM versus  WIMP}
Theoretically, the ADM provides a natural, simple, and predictive framework to
understand the dark matter relic density~\cite{Barr:1990ca,Kang:2011wb,Graesser:2011wi}.
This framework shows prominent advantages over the well accepted WIMP framework.
In the first, it resolves the cosmological coincidence problem or naturalness problem, i.e.,
it explains the origin of the energy density ratio between the visible and invisible matters
in the present University, ${\Omega_Bh^2}:{\Omega_{\rm DM}h^2}\sim 1:5$.
In the ADM framework, both the DM and baryon number densities are dominated
by their asymmetric components, and some mechanism is assumed to relate them: 
$n_D-n_{\bar D}=C(n_{B}-n_{\bar B})$ with $C$ a model dependent constant. 
So their relic density ratio is
   \begin{align}\label{ADM}
\f{\Omega_{\rm DM}h^2}{\Omega_Bh^2}=\f{m_D(n_D-n_{\bar D})}{m_p(n_B-n_{\bar B})}=C\times\f{m_D}{m_p}.
 \end{align}
Once $C\sim 1$~\footnote{Although many recent attempts generalize this constant to be a widely varying number in some models,
 it loses the prediction on the DM mass. In this paper, we stick to $C\sim 1$ by
claiming that the ADM obtains asymmetry via simple chemical equilibrium.} is determined,
the coincident puzzle is resolved given the single parameter, i.e., the mass ratio
$m_D/m_p$ is around 5. Hence a light DM about 5 GeV is predicted. By contrast, in the WIMP
framework their densities are independent on each other and then the observed ratio
is just a coincidence.

Next, in spite of the well-known miracle in predicting the correct order of relic density,
there exists the hidden fine-tuning for WIMP.
 But the ADM does not. The WIMP relic density is determined
by the freezing-out dynamics and typically scales as~\cite{WIMP}
 \begin{align}\label{}
\Omega_{\rm DM} h^2\propto \f{1}{\langle\sigma v\rangle}\sim \f{m_{\rm DM}^2}{g_{\rm DM}^4},
 \end{align}
with $g_{\rm DM}$ the DM and light particle effective coupling. This equation merely is
a schematic estimation, but it indicates that the WIMP relic density may be sensitive to
the fundamental parameters (higher powers of parameters). Even worse, the WIMP relic density,
for instance, of the well-tempered neutralino in the MSSM~\cite{ArkaniHamed:2006mb},
may hide a source of fine-tuning. In this regard, the ADM is more natural because
its relic density is irrespective of the concrete value of couplings. Additionally, the complicate
Boltzeman equations are no longer quantificationally important for the ADM relic density.

By construction, the ADM should carry a continuous (at least highly approximately)
conservative charge~\footnote{For details, please see a recent work on the charge breaking term
effect in the Ref.~\cite{Buckley:2011ye}.}. Consequently, the most popular DM candidates,
such as the neutralino, gravitino and real scalar can not be the ADM candidate since they are
CP self-conjugated particles, and thus they can not carry any continuous charge.
So, how to let them share the great merits of ADM is a very interesting question.
In this paper, we consider gravitino as a concrete example.

\subsection{The Gravitino Relic Density from an AMP}

Recall that the essential point of the ADM is that its number density is dynamically related to
the baryon number density, it is tempting to conceive a generalized ADM framework: the DM $\chi$, such as
the aforementioned DM candidates, is not necessary to be an ADM candidate, and an
asymmetric metastable particle (AMP), which has similar dynamic features as the ADM, transfers
its asymmetric number density to the actual DM density. 
In this way, the non-ADM candidate also enjoys the ADM merits.

This mechanism to account for the DM relic density is the combination of the ADM and
non-thermal dark matter production. So, to make our mechanism work, there are two basic
hypothesis inherited from them
\begin{itemize}
  \item $\chi$ has negligible  thermal relic density. If it is used to enter the plasma and freezes out
at the temperature $T_f\sim m_{\chi}/20$, its relic density should be small. Oppositely, if it never
undergoes the thermal equilibrium, such as the superWIMP gravitino $\wt G$, its yield after the inflation
should be suppressed.
  \item An AMP late decays to the DM at a time $\tau_{\rm AMP}$: For the superWIMP-like DM,
the time is not constrained directly except for some cosmological bounds;
While for the WIMP-like DM, $\tau_{\rm AMP}$ is required to be after the DM thermal decoupling, i.e.,
       $\tau_{\rm AMP}\gg H^{-1}(T_f)\simeq 0.3g_*^{-1/2} M_{Pl}/T_{f}^2$,
  which can be transferred to the bound on the AMP decay rate:
\begin{align}\label{AMP}
    \Gamma_{\rm  AMP}\ll 3.8\times10^{-18} \times \L\f{m_{\rm AMP}}{100{\rm GeV}}\R^2{\,\rm GeV}.
\end{align}
\end{itemize}
Comments on the new framework are in orders. First, the DM  can share
the merits of ADM. Next, the AMP mass is not constrained to lie around 5 GeV,
since on our purpose only its asymmetric number density $n_{\rm AMP}$ is of crucial importance.
Finally, for the bosonic ADM which does not annihilate today, the bounds from
neutron stars constraint on it stringently~\cite{ADMCONSTRAINT}. 
Oppositely, here the DM should have sufficiently large
annihilation rate, at least for the WIMP-like DM, and then the constraints are avoided.

Directly connecting to the SUSY breaking, the gravitino is a specially attractive and natural candidate
of this framework~\footnote{In some secluded or very weakly coupled MSSM dark sector, for example,
in the light $U(1)_X$ sector~\cite{Kang:2010mh}, the light DM particle typically annihilate effectively
when leaving with very  small relic  density. But they can acquire the asymmetric relic density
from the visible sector AMP.}. Within the supergravity, the gravitino mass is uniquely determined by
the hidden sector SUSY-breaking scale $\sqrt{F}$
  \begin{align}\label{}
  m_{\wt G}=\f{F}{\sqrt{3}M_{Pl}},
 \end{align}
with $M_{Pl}\simeq 2.14\times 10^{18}$ GeV the reduced Planck scale. $\sqrt{F}$ lies between
$10^{5}{\,\rm GeV}$ and $10^{11}$ GeV, depending on the scheme of SUSY breaking mediation.
Hence $m_{\wt G}$ varies widely, from the eV to TeV region. However, if $\wt G$
is the dominant DM component, it is possible to fix $m_{\wt G}$ which in turn
has deep implication to the SUSY breaking scale as well as its mediation mechanism.
In the following, we present some scenarios which can potentially determine or predict $m_{\wt G}$:
\begin{description}
  \item[Thermal gravitino] The original supersymmetric DM candidate is nothing but the gravitino, proposed by H. Pagels
and J. R. Primack~\cite{Pagels:1981ke}. They showed that for the thermal gravitino, its relic density
is
   \begin{align}\label{}
\Omega_{\wt G}^{th}h^2\simeq 0.1\L\f{m_{\wt G}}{100\rm\, \,eV}\R\L\f{106.75}{g_{*S,f}}\R,
\end{align}
where $g_{*S,f}$ is the relativistic degree of freedoms in the plasma. To produce
the correct relic density, $m_{\wt G}\sim100$ eV, which means a very low SUSY breaking scale,
is determined. Nevertheless, the astrophysical observations exclude such a hot DM~\cite{Lyman}.
Even stronger, the combination of the WMAP, CMB and Lyman-$\alpha$ data excludes the thermal
gravitino to make up of the whole DM, irrespective of its mass~\cite{Feng:2010ij}.
  \item[Non-thermal gravitino] If the gravitino never enters the thermal equilibrium, it can be
  a viable DM candidate with correct relic density by tuning the reheating temperature~\cite{G:lowreh}.
  For the non-thermal $\wt G$, its thermal relic density is via the scattering, and is
  linearly proportional to the reheating temperature $T_R$~\cite{Bolz:2000fu}
  \begin{align}\label{}
\Omega^{th}_{\wt G}h^2\simeq 0.03\L \f{10\rm\,\,GeV}{m_{\wt G}}\R \L\f{M_{3}}{3\rm\,\, TeV}\R^2\L\f{T_R}{10^6\rm\,\, GeV}\R,
 \end{align}
which is valid for $T_R<T_{f}$ with $T_f$ the thermal gravitino decoupling temperature.
But we have no data to trace back to the so early Universe and consequently $T_R$ is
unknown yet, rendering the unique prediction on the gravitino mass loss.
  \item [``Asymmetric" gravitino] We turn to our scenario. Assuming that the thermal
  gravitino production is ignorable due to either a rather low reheating temperature
  or relatively light gluino, then the non-thermal production from the AMP late decay leads to
  \begin{align}\label{mD}
m_{\wt G}=\f{\Omega_{\rm DM}h^2}{\Omega_Bh^2}\f{n_B}{n_{\rm AMP}}m_p.
 \end{align}
Because the AMP number density $n_{\rm AMP}$ is about the baryon density $n_B$,  
 the gravitino mass is predicted to be $\sim5$ GeV.
\end{description}

We would like to emphasize that, the ``asymmetric" gravitino scenario completely solves the dark matter naturalness problem.
Most ADM models are not thoroughly natural since they leave us with the question on the origin of the ADM mass scale.
Here \emph{the gravitino mass is dynamically generated and tied to the dynamical SUSY breaking
scale}. From this point of view,
the ``asymmetric" gravitino scenario should receive enough attention.

\subsection{Hybrid Gravity-Gauge Mediated Supersymmetry Breaking }

In this subsection, we simply assume that there is a successful AMP to account for
the gravitino relic density, and then study its implication to the supersymmetric models.
We defer the detailed discussion on the conditions of a viable AMP in the next subsection.

A gravitino mass of several GeVs has far-reaching implication to SUSY.
It implies that the SUSY-breaking should be mediated by the hybrid of gravity mediation and gauge mediation (in the anomaly
mediation the gravitno mass is around tens of TeVs). To see it, notice that the gravity mediation typically
contributions to soft terms at the order of gravitino mass
  \begin{align}\label{PMSB}
m_{\wt f}^2\simeq m_{\wt G}^2,\quad A\simeq M_{\ld}\simeq m_{\wt G}.
 \end{align}
where $m_{\wt f}$, $ M_{\ld}$ and $A$ are sfermion masses, gaugino masses, and trilinear soft terms,
respectively. Phenomenologically, it is required that the soft terms give
$m_{\wt f}\sim M_{\ld}\sim {\cal O}$ (500) GeV, which is much heavier than 5 GeV. Therefore, in the
``asymmetric" gravitino scenario the pure gravity mediation is not enough to account for the origin of
supersymmetry breaking soft terms. 
We have to turn to the GMSB~\cite{GMSBearly}, where the gravitino mass is
  \begin{align}\label{}
m_{\wt G}=\f{1}{\sqrt{3}}\f{\Ld M_{mess}}{M_{\rm Pl}}.
 \end{align}
The realistic soft spectrum requires $\Ld=F/M_{mess}\sim10^{4}-10^{5}$ GeV with $M_{mess}$ the messenger
mass scale. A 5 GeV gravitino implies $M_{mess}\sim 10^{14}$ GeV, close to
the grand unification scale.

In this hybrid mediation scenario, the flavor structure of soft terms coming from the GMSB
satisfies the minimal flavor violation (MFV) hypothesis. However, the anarchical
soft mass terms from the gravity mediation, in spite of merely the sub-dominant contribution,
reach flavor violations to the most extent. It implies that the ``asymmetric" gravitino scenario
is testable from the future  high-precision FCNC experiments in the B-factories and the LHCb
via  $c\ra u,\, b\ra d,s$, and $t\ra c,u$ decays,  and can be explored at the LHC~\cite{Hiller:2008sv}.

The decay $c\ra u\bar qq$ might have left tracks in the recent LHCb experiment. It reports
a measurement of the CP asymmetry in the $D$ meson decay, $A_{CP}(D^0\ra K^+K^-)-A_{CP}(D^0\ra \pi^+\pi^-)\neq0$,
with the measured value deviating from the SM predication by 3.5 $\sigma$ evidence~\cite{LHCb}.
This deviation can be interpreted by a new direct CP violation in the $D$ decay.
However, how to generate such a large CP violation in the decay meanwhile suppress
the CP violation in the $D^0-\bar D^0$ mixing is challenging~\cite{LHCb:papers}.
Our framework offers a solution and the points can be found in Ref.~\cite{DCPV}:
the up squark-gluino loop with the left-right mixing mass insertion $\delta_{LR}\equiv
(\wt m_{LR}^{2u})_{12}/{\wt m}^2$ ($\wt m$ is the common squark mass)
contributes both to the Wilson coefficients of the dipole operator which
contributes to the direct CP violation in the $D$ decay, and to the operator
generating the $D^0-\bar D^0$ mixing. The former is enhanced by $m_{\wt g}/m_c$
while the latter is not, thus the tension is avoided naturally.
We leave the quantitatively study elsewhere.

\subsection{The Scattering Induced Charge Wash-Out Effects}\label{washout:sc}
In SUSY the existence of an AMP is of great theoretical interest, but unfortunately it can not
be accommodated within the MSSM or its simple extension.
To show that, we consider the AMP system with particle (1) and anti-particle (2), and their
individual number densities can change via the following annihilation and scattering processes
\begin{align}\label{}
12,21\leftrightarrow f\bar f,\quad 11\leftrightarrow ff,\quad 22\leftrightarrow \bar f\bar f,\quad 1\bar f\leftrightarrow 2f,
  \end{align}
with $f$ the SM particle. Such a simplified system encodes the main features of the number density
evolutions of the AMP candidates within the MSSM, for instance, the sfermions and charginos.
The first reaction conserves a charge of the sparticle (we call it sparticle charge for short),
while the others do not and will be proven to render the AMP impossible.

For the proof, we systematically consider the charge evolution through Boltzmann equations.
The number density with the chemical potential factored out is defined as
  \begin{align}\label{}
n_1^{eq}=a z^{3/2}e^{-z}e^{\xi_1(z)},\quad
n_2^{eq}=a z^{3/2}e^{-z-\Delta m/T}e^{\xi_2(z)},
  \end{align}
where $a\equiv0.064gT^3$ with $g$ the particle internal degree of freedom. Also,
$z\equiv m/T$ and $\xi_i\equiv\mu_i/T$, where $m$ is the particle mass and $\mu_i$
are the particle chemical potentials. Here we have the mass difference $\Delta m=0$.
With the help of this notation, the Boltzmann equation can be written in the following form
  \begin{align}\label{BE1}
-zy_1'-zy_1\f{Y_{eq}'}{Y_{eq}}=&(y_1y_2-1)\frac{\Gamma_{12}}{H}+(y_1^2-e^{-2\xi_f})\frac{\Gamma_{11}}{H}+(y_1e^{-\xi_f}-y_2e^{\xi_f})\frac{\Gamma_{1\bar f}}{H},\\
-zy_2'-zy_2\f{Y_{eq}'}{Y_{eq}}=&(y_1y_2-1)\frac{\Gamma_{12}}{H}+(y_2^2-e^{2\xi_f})\frac{\Gamma_{22}}{H}+(y_2e^{\xi_f}-y_1e^{-\xi_f})\frac{\Gamma_{2  f}}{H},
\label{BE2}
  \end{align}
where $H$ is the Hubble constant, $y\equiv Y/Y_{eq}$,
$\Gamma_{ij}=\gamma_{ij}/n_{eq}$, and ${Y_{eq}'}/{Y_{eq}}=3/2z-1\simeq -1$. Approximately,
the reaction density of the reaction $ij\ra ab$ is $\gamma_{ij}\approx n_i^{eq}n_j^{eq}\langle\sigma v\rangle_{ij\ra ab}$.
When the reaction rate $\Gamma_{ij}\gg H$, the corresponding reaction enters the chemical equilibrium, and is ignorable in the opposite.
The scatterings between the AMP and SM particles do not change the total number density
of the system, but convert the particle and antiparticle to each other. In the following, we will show that
it is highly relevant to the asymmetry of the system.

During the AMP cosmological evolution, when it enters the deep non-relativistic region, its number density
is power suppressed and the highly suppressed annihilation rate thereof. However, the scattering rate is
much less suppressed than the annihilation rate, manifested in the following estimation
\begin{align}\label{VCratio}
\f{\Gamma_{anni}}{\Gamma_{scat}}\sim \f{(n_1^{eq})^2\langle\sigma v\rangle}{(n_1^{eq})\langle
\sigma v\rangle}\sim n_1^{eq}\sim z^{3/2}e^{-z}~.
  \end{align}
To get the final result, we have reasonably assumed that there is no great disparity between the scattering and
annihilation cross sections. Then at the (WIMP-like) AMP decoupling temperature $z\sim z_f\simeq 25$
the annihilation rate is smaller than the scattering rate by 9 orders. This phenomena can be understood by
nothing but that during the WIMP decoupling the kinetic decoupling happens much
later than the chemical decoupling~\cite{Griest:1990kh}. In other words, even long after the annihilation
freezing-out, the scattering still maintains the chemical equilibrium between 1, 2 and $f,\bar f$, as establishes the
following relation:
\begin{align}\label{eq}
\xi_1-\xi_f=\xi_2+\xi_f.
  \end{align}
Note that in the plasma the asymmetry of the relativistic SM particles are tiny, at the order of
the baryon asymmetry, so we get
\begin{align}\label{NRCP}
\f{n_f-n_{\bar f}}{n_{\gamma}}\sim
\f{n_b-n_{\bar b}}{n_{\gamma}}=\eta\Rightarrow \f{g_f}{2}\f{\pi^2}{3}\xi_f\sim\eta\sim10^{-10}.
  \end{align}
Now, we draw the conclusion that the AMP asymmetry is ignorably small.

For convenience, we give the condition of decoupling the scattering process.
The particle-antiparticle number density ratio is
\begin{align}\label{}
r_{eq}=\L \f{n_1^{eq}}{n_2^{eq}}\R=e^{\xi_1(z)-\xi_2(z)}.
  \end{align}
Thus, to guarantee the scattering freezing-out before the annihilation freezing-out, we have to suppress the
scattering cross section by orders of magnitude. For example, if the scattering process decouples
at the weak scale, it is required
  \begin{align}\label{suppress}
T^3\sigma_{sc}<H(T)=1.66g_*^{1/2}T^2/M_{\rm Pl}|_{T=100\rm\,GeV}~.
  \end{align}
From it we get $\sigma_{sc}<10^{-10}$ pb, about 10 orders smaller than the typical WIMP
annihilation cross section. Without special treatment, such a great suppression is impossible
in the simple models like the MSSM.

In the claim of no AMP in the MSSM, there is a loop-hole. It is supposed that
as the sfermion $\wt f$ becomes non-relativistic, its SM partner $f$ is still
relativistic, but it is not always the case. If the mass of $f$ is close or even
heavier than the mass of $\wt f$, they decouple almost simultaneously. But now
$f$ is non-relativistic and has the maximal asymmetry:
  \begin{align}\label{NRCP1}
0.3g_fz^{3/2}e^{-z+\xi_f}\L1-e^{-2\xi_f}\R\sim \eta\Rightarrow 0< \xi_f\simeq z-10~.
  \end{align}
Thus, Eq.~(\ref{eq}) does not mean that $\wt f$ has an extremely small asymmetry. We consider
two examples: (I) The chargino system $(\wt W^\pm,H_{u}^+,\wt H_d^-)$ where
the vector bosons $W^\pm$ and charged Higgs bosons $H^\pm$ have
comparable masses to their superpartners.
However, $W^\pm$ and $H^\pm$ mediate scattering processes such as $\wt C^\pm_1+u\ra \wt N_1+d$, where
$\wt C_1,\,\wt N_1$ are the lightest chargino and neutralino respectively. Even if
$M_{\wt C_1}<M_{\wt N_1}$ is obtained in some region of the MSSM parameter space,
they are still quite degenerate. Consequently the scattering can proceed fast enough to
establish the chemical equilibrium between the charigno and light quarks, which leads to
$\xi_{\wt C_1}=\xi_{d_L}-\xi_{u_L}$, and therefore no asymmetry for the charigno is left.
(II) The stop-top system where the lighter stop $\wt t_1$ can be lighter than
the top quark in some cases. However, even we do not consider the effective
$\wt t_1 t\ra \wt t_1^*\bar t$ at the tail of the Boltzmann distribution, the scattering reaction
$\wt t_1 c\ra \wt t_1^*\bar c$ via the CKM mixing can only be suppressed by $\ld^8\sim 10^{-7}$
with $\ld\simeq0.22$ the Cabbibo angle, which is several orders short to decouple the scattering.
In light of the above arguments, we safely  draw the conclusion:  in the MSSM no AMP can be accommodated.

\begin{figure}[htb]
\begin{center}
\includegraphics[width=5.5in]{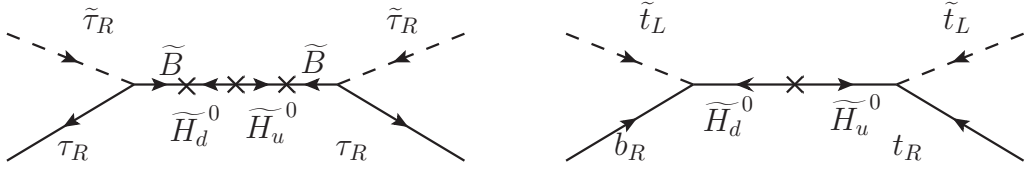}
\end{center}
\caption{The gaugino and Higgsino mediated sparticle charge washing-out processes.
}
\label{SCVwash}
\end{figure}
To end up this subsection, we would like to add a comment that the absence of AMP within the MSSM
has a close relation with the $U(1)_R$ symmetry.
This continuous symmetry does not communicate with SUSY, and in the superspace it is defined as
(in the following $W$ is the superpotential, and $\theta$ is the Grassmanian coordinate)
    \begin{align}\label{}
(d\bar\theta)\,\theta\ra e^{i\alpha}(d\bar\theta)\theta,\quad (d\theta)\bar\theta\ra e^{-i\alpha}(d\theta)\bar\theta, \quad {W}\ra  e^{2i\alpha} W.
  \end{align}
Then for a chiral superfield $\Phi$ with $U(1)_R$ charge $r$, its lowest and $\theta-$component carry
$U(1)_R$ charge $r$ and $r-1$, respectively. In particular, gauginos must carry $U(1)_R$ charge $-1$.
Fig.~\ref{SCVwash} shows the typical processes which wash out the sparticle charge. We may describe them
through the effective operator ${\cal O}_{SCV}=\wt f\wt f\bar f \bar f$, which breaks the $U(1)_R$ charge by $-2$.
It has two origins: one is from gauge interactions,
generated with insertions of the Majorana gaugino mass terms; The other one is from the Yukawa coupling terms
in the superpotential with insertions or the $\mu-$term. And their contributions to such operators are always there.
But their strengths can be controlled by Yukawa couplings (by accidently, in the exact $U(1)_R$ MSSM,
the $\mu-$term will be forbidden out of other consideration, and we shall return to it later).


\section{The AMP in the MSSM Extensions }

In light of the general lessons from the previous Section, to construct a supersymmetric model with an AMP,
one may properly introduce extra vector-like matters to the MSSM. Note that such extensions have been investigated from
some other motivations. On top of that, based on the symmetry analysis, the $U(1)_R-$symmetric MSSM (MRSSM)
also provides an interesting extension.

\subsection{The AMP Inspired  vMSSM}

The MSSM extended with weak-scale vector-like  particles (vMSSM) arises in the various model building~\cite{Barger:2006fm, Liu:2009cc},
for example, in the stringy models of particle physics~\cite{Barger:2006fm} or in the MSSM extending with an
anomaly-free group such as $U(1)_R$~\cite{Evans:2011uq}. The presence of such light particles at the low energy
brings rich and prominent phenomenology, e.g., lifting the lightest CP-even Higgs boson
mass~\cite{Martin:2009bg} and accordingly solving the little hierarchy problem~\cite{Babu:2008ge, Graham:2009gy}.
In this work, we primarily investigate the accommodation of an AMP.

 The vMSSM has  different versions depending on what vector-like matters are introduced. We require
that the particle content is consistent with GUTs, for example, the $SU(5)-$GUT.
As a minimal attempt, we consider the MSSM+$\bar 5_4+5_4+\bar N^c+N^c$ with the ordinary matter
family indice denoted by $i=1,2,3$. Decomposed into the SM,
${\bar 5}_4=(D_4^c,L_4)$, where the SM gauge group quantum numbers for the components are
$D_4^c=(3,1,-1/3)$ and $L_4=(1,2,-1/2)$. Also, $N^c$ and $\bar N^c$ are SM singlets.
To not be excluded, the vector-like pairs should have large supersymmetric mass terms. 
The superpotential in our model is
  \begin{align}\label{vMSSM}
W_{Z_2}\supset &\L y_{ij}^UQ_iH_uU_j^c+ y_{ij}^DQ_iH_dD_j^c+y_{ij}^E L_iH_dE_j^c\R+\L k_NH_uL_4 N^c-h_NH_d\bar L_4 \bar N^c\R,\cr
&+\L M_DD_4^c \bar D_4^c+M_LL_4\bar L_4+M_NN^c\bar N^c\R,\cr
-V_{soft}\supset &\sum_{\phi_i=D,\bar D,...} m_{\phi_i}^2|\phi_i|^2 +\L B_{D}D_4^c\bar D_4^c+B_LL_4\bar L_4+B_NN^c\bar N^c+h.c.\R\cr
&+\L A_kH_uL_4N^c-A_hH_d\bar L_4\bar N^c+h.c.\R.
  \end{align}
The implication of the subscript $Z_2$ will be clear. Obviously, in Eq.~(\ref{vMSSM}) the terms
involving the vector-like particles respect a global $U(1)$ symmetry, so we can expect
an AMP candidate. Among these terms, $k_NH_u L_4N^c$ with $k_N\sim1$ is of particular importance,
since it not only helps to increase the Higgs boson mass but also is crucial for the realization of the AMP.

We focus on the case where the singlet  $N^c$ and $\bar N^c$ sector provides 
the AMP candidate. As a working simplification, one
can decouple the SM-charged vector-like particles $(D^c_4, {\bar D_4}^c)$ and $(L_4,\bar L_4)$ by setting
rather large vector-like masses: $M_D\simeq M_L\sim{\cal O}$ (TeV). Note that in the GMSB their superpartners
acquire large positive soft mass squares, while the soft bilinear terms $B_{D,L,N}$, which
may reduce the soft masses, are generated only via the renormalization group equation (RGE)
effect and thus are suppressed. In summary, their superpartners are even heavier than the vector-like fermions.

Now we turn our attention to the singlet sector. Due to the mild hierarchy $k_Nv_u\sim M_N\ll M_L$,
we can ignore the mixings between the singlets and the neutral components of the doublets. As a result,
$\bar\nu_{4}$ pairs with $\nu_4$ and gets a large Dirac mass $\simeq M_L$. As a result, $\bar N^c$ and $N^c$
form an isolated singlet system, of which the scalar singlets have mass squares:
\begin{align}\label{}
m_{\wt {N}_1}^2\simeq M_N^2+m_{\wt { N^c}}^2, \quad m_{\wt { N}_2}^2\simeq M_N^2+m_{\wt {\bar N^c}}^2~.~
\end{align}
$B_N$ has been set to be zero as discussed previously. Since
$\wt{ N^c}$ and $\wt {\bar N^c}$ are singlets, their soft mass terms vanish at the UV boundary and
are generated by the RGE effects via their couplings to the doublets, for example,
\begin{align}\label{}
m_{\wt { N^c}}^2\sim -\L\f{k_N^2}{16\pi^2}\log\f{M_{\rm GUT}}{M_N}\R m_{\wt L_4}^2~.
\end{align}
It is negative and moreover large due to the order one $k_N$. Similar estimation is applied to
$m_{\wt {\bar N^c}}^2$. We assume $h_N\ll k_N$, which is a reasonable relation to keep the perturbativity, taking
$k_N\sim1$ into account. Hence we have $-m_{\wt {N^c}}^2\gg -m_{\wt {\bar N^c}}^2$ and $\wt{N^c}\approx\wt{N}_1$ is the AMP.

The $k_N$ term also provides a sufficiently effective channel to let the AMP annihilate away the symmetric
part. For example, the AMP may annihilate into a pair of Higgs bosons via the contact term
$|F_L|^2\supset k_N^2 |\wt{N^c}|^2 h^2/2$, with a thermally averaged cross section estimated to be
\begin{align}\label{}
\sigma v\sim \f{1}{8}\f{k_N^4}{32\pi}\f{1}{m_{\wt N_1}^2}\sim 10\rm\,pb,
\end{align}
where we have taken $m_{\wt N_1}\simeq 100$ GeV. This cross section is large enough to annihilate away
the AMP symmetric component~\cite{Graesser:2011wi}. Note that if $\wt N_1$ is lighter than the
SM-like Higgs, one can still gets a large cross section by turning to the $s-$channel resonant annihilation
$\wt N_1\wt N_1^*\ra W^+W^-/\bar ff$, exchanging the heavier Higgs or CP-odd Higgs.

The model has some phenomenological demerits and requires improvements. At first, the vector-like
mass terms are introduced  by hand, but they may share the common origin as the $\mu$ parameter in
the MSSM. The weak scale $\mu$ can be produced in the presence of an intermediate scale like the
$U(1)_{PQ}$ spontaneously breaking scale, which falls into a window $f_{PQ}\sim10^{9}-10^{12}$ GeV.
Then, the supersymmetric mass terms are generated via the high-dimension operators as follows~\cite{PQmu}
 \begin{align}\label{SU5}
 \f{X^2}{M_{\rm Pl}}\L \ld_4 5_4\bar 5_4+ \ld_h5_u\bar 5_d+\ld_11_4\bar1_4\R,
  \end{align}
where $\ld_4$, $\ld_h$ and $\ld_1$ are Yukawa couplings, and 
$X$ is the PQ-axion superfield carrying PQ-charge normalized $-1$, and it develops
a vacuum expectation value around $f_{PQ}$. So we get the mass scales as
\begin{align}\label{}
\mu=\ld_h f_{PQ}^2/M_{\rm Pl}, \quad M_4=\f{\ld_4}{\ld_h}\mu,\quad M_1 =\f{\ld_1}{\ld_h}\mu.
  \end{align}
The operator coefficients $\ld_{4,h}$ are order one, but $\ld_1$ is moderately smaller for the sake of
a smaller $M_N=M_1$.

Second, this vMSSM conserves an exact $Z_2-$parity, also the global $U(1)$ addressed before,
that acts only on the exotic particles. This symmetry forbids the mixings between vector-like
particles and MSSM matters. Consequently, the model suffers
from an acute cosmological problem, i.e., the (quasi) stable colored particles with significant relics
are definitely excluded by observations. Even worse, it renders the ambitious ``asymmetric"
gravitino scenario to be stillborn. Firstly, the lightest vector-like particle $\wt N_1$
is stable rather than decay. Secondly, for an exact $Z_2$ the dark number of $U(1)$ is conserved,
so no final asymmetry can be generated during the Universe evolution, except the genesis of
primeval dark number. In this paper we only consider the simple case that \emph{the dark sector asymmetry is
obtained by means of the re-distribution of the visible sector asymmetry}.

So it is necessary to introduce $Z_2-$breaking terms, and then the $R-$parity is
the unique low energy symmetry. The soft $Z_2-$breaking terms under consideration are
  \begin{align}\label{mixing:4i}
W_{\K}\supset &\epsilon_{i}^uH_d Q_iU_4^c+\epsilon_{i}^eH_d L_4E_i^c+\epsilon_{i}^NH_u L_i N^c.
  \end{align}
We require $\epsilon_i\ll1$, consistent with avoiding large flavor violations, and the
smallness may be due to some flavor symmetry which addresses the hierarchy structure of the MSSM Yukawa couplings.
Note that these parameters, together with the vector-like particle masses,
do not actually enlarge the MSSM parameter space, since their concrete values are not of crucial
importance. Such properties reflect the naturalness of the AMP, and justify
our frequently using of estimation rather than precise calculation.

 Alternatively, one can introduce the vector-like particles $(10_4,\bar 10_4)$
 with $10_4=(Q_4,U_4^c,E_4^c)$, and the discussion proceeds similarly. They couple to the matter fields via
the following superpotential
  \begin{align}\label{}
W_{V,10}\supset& 10_a10_b5_u+10_a\bar 5_i\bar 5_d+\bar 10_4\bar 10_4\bar 5_d+M_{10}10_4\bar 10_4\cr
\supset&  Q_aU_a^cH_u+Q_aD_i^cH_d+Q_aE_i^cH_d+\bar Q_4\bar U_4^cH_d~,
  \end{align}
where  some terms may have to be forbidden by the PQ-symmetry.
 In the GMSB    $(\wt E_4^c,\wt {\bar E}_4)$ are lighter than the colored components, but they are heavier than the pure right-handed sleptons $\wt\tau_R$ in the MSSM and then not the NLSP.
The GMSB contribution gives $m_{\wt \tau_R}^2\simeq m_{\wt E_4^c}^2 \simeq m_{\wt{\bar E}_4^c}^2$, but $(\wt E_4^c,\wt {\bar E}_4)$
  have  large vector-like mass term,
  while the splitting due to $B_{E}$ is loop suppressed and ignorable, so quite generically they are
   heavier. However, in more complicated situations the extra colored sparticles can be the NLSP.
  For example,  coupling $H_u$ to the messengers can lower the soft masses $m_{\wt {\bar Q}_4}^2$
 and $m_{\wt {\bar U}_4^c}^2$, or similarly we can solve the
 $\mu/B_\mu$ problem in the GMSB by coupling them directly to the messenger sector~\cite{Kang:2011az}.
 Thus, one may get more general spectra and allow for extra squark being the NLSP.

To end this subsection, we give a discussion on the features of the AMP late decay and the subsequent
impact on the cosmology. Since we are working in the hybrid gravity-gauge mediated SUSY-breaking framework,
in which the LSP is the gravitino with mass around 5 GeV and the NLSP is the AMP $\wt N_1$ with mass about 100 GeV,
the $R-$parity odd AMP must dominantly decay to the gravitino plus neutrinos via the two-body 
decay $\wt N_1\ra \nu_i+\wt G$.
The decay width is highly suppressed~\cite{Martin:1997ns}:
\begin{align}\label{}
\Gamma(\wt N_1\ra \nu_i+\wt G)&\approx\L\epsilon_i^Nv_u/M_N\R^2 \f{m_{\wt N_1}^5}{48\pi M_{pl}^2m_{\wt G}^2}\cr
&\simeq0.4\times10^{-11}\L\f{\epsilon_i^Nv_u/M_N}{10^{-3}}\R^2\L \f{m_{\wt N_1}}{100\,\rm GeV}\R^5\L \f{5\,\rm GeV}{m_{\wt G}}\R^2s^{-1},
  \end{align}
the first factor in the first line denotes the mixing between $N^c$ and $\nu_i$. Therefore, the decay
happens well after the BBN, and more exactly at the time the formation of structure begins.
The AMP decay is cosmologically safe based on the following facts: (A) The electromagnetic (EM) BBN
constraint on the sneutrino-like NLSP late decay is rather weak for the decay time $t>10^7$ s; (B) The hadronic decay modes
at the three-body level $\wt N_1\ra \nu_i Z\wt G,\ \ell_iW\wt G$ are completely free of the
hadronic BBN constraints~\cite{Feng:2004mt}. The only property of the AMP decay which may receive
cosmological interest is that the warm gravitino may help to reduce the power spectrum on the small scale~\cite{WDM}.

\subsection{The AMP Asymmetry }

In the plasma at a temperature $T$, if a particle $\phi$ is in the thermal equilibrium and
has a chemical potential $\mu_\phi$, its asymmetry is given by
\begin{align}\label{as-30}
n_{+}^{eq}-n_{-}^{eq}=&f_{b,f}(m/T)\times\f{g T^2}{6}\mu_\phi,\cr
 f_{b,f}(m/T)=&\f{6}{\pi^2}\int_0^\infty dx
\f{x^2\exp[-\sqrt{x^2+(m/T)^2}]}{\L\theta_{b,f}+\exp[-\sqrt{x^2+(m/T)^2}]\R^2},
  \end{align}
where $\theta_{b,f}=\mp1$ are for a boson and a fermion respectively. The function $f_{b,f}(m/T)$ denotes the
particle threshold effect. In the relativistic limit $m\ll T$, we get $f_{b,f}\ra 2,$ 1 respectively. Thus,
they recover the usual linear relations between the chemical potentials and the number densities.
Oppositely, \emph{when the thermal $\phi$ enters the non-relativistic limit,
$f\ra0$,  $\phi$ has ignorable asymmetry} as expected.

The precise determinations on the final asymmetries in dark and invisible sectors involve several
temperatures: the electroweak (EW) sphaleron decoupling temperature $T_{sp}$, the EW phase 
transition critical temperature $T_c$ and $T_D$. 
Below $T_D$ the interactions maintaining the chemical equilibrium between the AMP and the SM fermions decouple. 
The relative magnitudes between $T_{sp}$ and $T_c$ suffer uncertainty, and we
may take $T_{sp}\lesssim T_c\sim100$ GeV~\cite{Davidson:2008bu}. Quantitatively, the other choice will give
similar result. Owing to the complicated interactions, determining $T_D$ is involved.
However, we are allowed to consider a simplified case which is consistent with the previous setup:
all the SM-charged vector-like particles are rather heavy and $\epsilon_{i}^{u,d}$ are irrelevantly small.
Thus, the terms of cosmological concern are reduced to
\begin{align}\label{equili}
W_{\rm vMSSM}'=\epsilon_{i}^NH_u L_iN^c+M_NN^c\bar N^c+W_{\rm MSSM},
  \end{align}
with the corresponding soft terms implied. We will find that the first term establishes the
required chemical equilibrium and converts the lepton number $L$ to
the singlets. For not extremely small $\epsilon_i^N$, the chemical equilibrium 
breaks down below the AMP mass scale, namely $T_D\sim m_{\rm AMP}$.

We now have to check that the asymmetry stored in the AMP $\wt N_1$ avoids the washing-out.
The charge washing-out scattering $\wt N_1\nu_i\ra \wt N_1^* \bar \nu_i$ comes
from the interactions $\ld_{il}\, \chi_l\nu_i\wt N_1+c.c.$ with
$\ld_{il}\approx\epsilon_i^NF_{l\wt H^0_{u}}+\sqrt{2}g_2F_{\wt\nu_i\wt N_1}$, in which $F_{l\wt H^0_{u}}$ and
$F_{\wt\nu_i\wt N_1}$ denote for the fractions of gauginos $\chi_l$ in $\wt H_u^0$ and
$\wt N_1$ in $\wt\nu_i$, respectively. At a low temperature, the thermally averaged
scattering cross sections are roughly given by
\begin{align}\label{wash}
\langle\sigma(\wt N_1\nu_i\ra \wt N_1^* \bar \nu_i) v\rangle\sim  \f{\ld_{il}^4}{4\pi}\f{1}{M_{\chi_l}^2}.
\end{align}
For the gaugino masses $M_{\chi_l}\sim 300$ GeV $ >m_{\wt N_1}$,
the cross sections can be suppressed to the order required in Eq.~(\ref{suppress})
when $\ld_{il}<{\cal O}(10^{-3})$. The Yukawa coupling contributions to $\ld_{il}$
are well under control as long as $\epsilon_i^N\lesssim10^{-3}$, but the gauge contributions which
arise from the mixing between the sneutrinos and $\wt N^c$ may be problematic. At the leading order, such mixings are due to
the universal gravity-mediated trilinear soft terms
\begin{align}\label{A:wash}
-{\cal L}_{soft}\supset A_{iN}\wt L_iH_u\wt N^c+c.c.,
\end{align}
with $A_{iN}\sim m_{\wt G}$, see Eq.~(\ref{PMSB}). In the case of $m_{\wt L_i}^2\gg m_{\wt N^c}^2\gg A_{iN}v_u$ we
get $F_{\wt\nu_i\wt N_1}\simeq A_{iN}v_u/m_{\wt L_i}^2$, so for a GeV-scale $m_{\wt G}$ the
suppression barely meets the previous upper bound.

We are now at the position to determine the chemical potential of particles in the vMSSM.
This is can be done via the standard method~\cite{Harvey:1990qw}:
when a reaction $A+B+...\leftrightarrow C+D+...$ has a rate faster than the
Universe expansion rate $H(T)$, we get an equation $\mu_A+\mu_B+...=\mu_C+\mu_D+...$, otherwise
a conservation number. Then the particle chemical potentials can be determined by solving all
these equations. Most of the calculations are similar to these in the ordinary MSSM~\cite{Dreiner:1992vm},
and the difference in the vMSSM will be commented on if it is necessary:
\begin{itemize}
  \item Due to the gauge interactions with the $SU(2)_L$ gauge boson $W^\pm$, the up and down components
in a $SU(2)_L$ doublet are in equilibrium:
  \begin{align}\label{W}
{[\rm down]}_L={[\rm up]}_L+W^-,\quad W^-=-W^+.
\end{align}
Hereafter we denote the particle name and its chemical potential with the same capital letter.
 \item The gaugino Majorana mass terms and the Higgs mixing term ($B_\mu-$term) give
    \begin{align}\label{soft}
 &\wt B=\wt W^0=0,\quad \wt W^+=-\wt W^-,\quad h_u^+=-h_d^-.
\end{align}
where $\wt W$ and $\wt B$ are respectively Wino and Bino, and $h_{u}^+$ and $h_d^-$ are charged Higgs bosons.
  \item By virtue of the gauge interactions with neutral gauginos,
particles and their superpartners have the same chemical potentials
\begin{align}\label{bino}
 & \wt W^-=W^-+\wt B=W^-(=-\wt W^+),\cr
&\wt f_L=f_L+\wt B=f_L,\quad \wt f_R=f_R-\wt B=f_R,\quad  \wt h=h-\wt B=h,
\end{align}
where $h$ denotes any Higgs doublet component, and $f_{L,R}$ are the chiral fermions.
These equations imply that the gauginos render the asymmetry stored in the sparticle to be
washed out.

   \item The Yukawa interactions mediated by neutral Higgs bosons equilibrate the chemical potentials of the
    left-handed and right-handed fermions.  
    \begin{align}\label{h0}
u_{R}=u_{L}+h^0_u,\quad d_{R}=d_{L}+h^0_d,\quad e_{R}=e_{L}+h^0_d,
\end{align}
All the three families of the SM fermions share the same chemical potential by virtue of the
fast family-exchanging reactions. Since we are considering $T_{sp}<T_c$, 
 the Higgs condensations means $h_u^0=h_d^0=0$ and then 
the left- and right-handed fermions have equal chemical potentials. 

\item We now make a special analysis on the singlet sector. 
The Yukawa interactions which keep the SM-singlet
    $N^c$ ($\sim N_R^\dagger$) in the chemical equilibrium with the SM fermions are $t_R+\ell_i\ra N_R+q_3$,
     whose rates are estimated as
\begin{align}\label{rate}
\Gamma_{ N_R}\sim3.8\times 10^{-3}{h_t^2(\epsilon_i^N)^2}T.
  \end{align}
  To derive it we have summed over the color and $SU(2)_L$ indices, and we are working in the massless
  limit. Thus, when $T\lesssim10^{8}\L\epsilon_i^N/10^{-3}\R^2$ GeV, $N_R$ enters the plasma and establishes the 
  chemical equilibrium with the left-handed neutrinos:
  \begin{align}\label{h0}
 N_R=\nu_L.\quad
\end{align}
Additionally, the vector-like mass term leads to $\bar N^c=-N^c=N_R$. 
Also, the scalar singlets enter the plasma through the Yukawa interactions
 in Eq.~(\ref{equili}) or its soft terms Eq.~(\ref{A:wash}). Either one leads to
     \begin{align}\label{}
 \wt {N^c}(=\wt N_R^\dagger)=-\wt{\bar N^c}= -\nu_{L}.
\end{align}
In summary the chemical potentials of singlets can be expressed with  $\nu_L$.
 \item Finally, the EW sphaleron process leads that the left-handed fermions satisfy
    \begin{align}\label{sph}
 &3(u_L+2d_L+\nu_L)=0\Leftrightarrow 3u_L+2W^-+\nu_L=0.
\end{align}
The vector-like doublets do not contribute to this equation since they do not contribute to the global $U(1)_{B/L}$ anomaly.
\end{itemize}
Eventually, four remained independent chemical potentials are for $W^-,\,u_L,~{\rm and}~\nu_L$.
A further constraint comes from the electromagnetic charge neutrality $Q=0$
\begin{align}\label{as}
Q
=&\L6u_L-6\nu_L-16W^-\R\f{T_{sp}^2}{6}=0~,
  \end{align}
where we do not include the heavy charged Higgs boson contributions.
And then we can solve Eqs.~(\ref{sph}) and (\ref{as}) and get
      \begin{align}\label{}
\nu_L=-\f{3(5+3f_{\wt e_R})}{1-3f_{\wt e_R}}u_L,\quad W^-=\f{3(2+3f_{\wt e_R})}{1-3f_{\wt e_R}}u_L.
\end{align}
Here $f_{\wt e_R}$ is the function defined in Eq.~(\ref{as-30}) with $m=m_{\wt e_R}$. Similar
notation is used for other particles.

Now we can determine the asymmetries of the baryon and AMP.
At first, the final baryon number density (only including the SM quarks) is
    \begin{align}\label{soft}
n_B=6\L 2u_L+W^-\R\f{T^2_{sp}}{6}=\f{6(8+3f_{\wt e_R})}{1-3f_{\wt e_R}}u_L\times\f{T^2_{sp}}{6}.
\end{align}
The conserved charge of the singlets is the lepton number, or exactly speaking, 
the sneutrino-number denoted as $\wt \nu$. 
But as argued before the ordinary sneutrino can not remain asymmetry due to the washing-out
effects during decoupling (this part will translate into the SM lepton number rather
than the sneutrino number), and only the part stored in the scalar singlets can survive:
   \begin{align}\label{}
n_{\wt {N^c}}=f_{\wt {N^c}}\times\wt\nu\f{T^2_{sp}}{6}=-\f{6(5+3f_{\wt e_R})}{1-3f_{\wt e_R}}u_L\times \f{T^2_{sp}}{6},
\end{align}
where we have taken $f_{\wt {N^c}}\sim2$. In turn, in light of the Eq.~(\ref{mD}) the gravitino mass is predicted to be
  \begin{align}\label{}
m_{\wt G}=\f{\Omega_{\rm DM}h^2}{\Omega_Bh^2}\f{8+3f_{\wt e_R}}{5+3f_{\wt e_R}}m_p\simeq 7.5\rm\,GeV,
 \end{align}
which does not depend on $f_{\wt e_R}<1$ much. If the contribution from $\wt N_2$ is comparable, the above prediction should 
be half.



\subsection{Remarks on  $U(1)_R$-Symmetric MSSM}

According to the analysis  in Section~\ref{washout:sc}, the gaugino/Higgsino
mediated wash-out effects originate from $U(1)_R-$symmetry breaking in the MSSM.
So, to forbid them, we are forced to turn to the $U(1)_R-$symmetric MSSM (MRSSM)~\cite{MRSSM}.
Surprisingly, although starting from  different motivations, we reach
the same picture: $U(1)_R-$symmetric GMSB.

The accommodation of AMP is  simple in the MRSSM, so we just make a shot comment in this paper.
In the MSSM, asides from the Majorana gaugino mass breaks
$U(1)_R$, so does $B_\mu-$term. And then it is not enough to introduce
 the Dirac partner of gauginos~\cite{muDirac}, {\it i.e.}, the
partners are $U(1)_R$ neutral adjoint  chiral superfileds under SM gauge groups
 $A=a+\theta \psi_A$. We have to extend
the Higgs sector by extra Higgs doublets with $R-$charge 2~\cite{MRSSM}, namely the $(R_u,R_d)$.
Thus, the MRSSM Higgs sector is
  \begin{align}\label{}
W_{MRSSM}\supset\mu_uH_uR_u+\mu_dR_dH_d~.~
\end{align}
The Yukawa couplings are identical with those in the MSSM. Technically,
the Higgsino mediated washing-out processes are forbidden  by the separation between
the $(H_u,R_u)$ and $(H_d,R_d)$.

We have to stress that the vanishing Higgsino-mediated contribution can not be
attributed totally to the exact $U(1)_R$. To see this, we consider the  MSSM without
$B_\mu-$term. The wash-out processes such as  $\wt f_L  f_R^*\ra \wt f_L^{'*} f'_R$ can proceed
(with rates suppressed by Yukawa couplings). But the successful  EW symmetry
breaking excludes such a scenario. Thus, the
exact $U(1)_R$ can only be realistic in the MRSSM.

\section{Conclusion and Discussion}

 We considered the hybrid gravity-gauge mediated supersymmetry breaking where
the gravitino mass is about several GeV. Interestingly,
the strong constraints on the supersymmetry viable parameter space
from the CMS and ATLAS experiments at the LHC can be relaxed due to the
heavy squarks and gluinos, and it is consistent with null
results in the DM direct search experiments such as
XENON100. Especially, the possible maximal flavor and CP violations
from the relatively small gravity mediation may naturally account for
the recent LHCb anomaly. In addition, because the gravitino mass is
around the asymmetric DM mass, we proposed the asymmetric origin of
the gravitino relic density and solved the coincident problem on
the DM and baryon densities $\Omega_{\rm DM}:\Omega_{B}\approx 5:1$.
The gravitino relic density arises from asymmetric metastable particle
(AMP) late decay. However, we showed that there is no AMP candidate in the
MSSM due to the robust gaugino/Higgsino
mediated wash-out effects. Interestingly, AMP can be realized in the well
motivated supersymmetric SMs with vector-like particles or continuous
$U(1)_R$ symmetry.

Some open question can be explored further to realize the ``asymmetric" gravitino framework.
For example, we can not exclude  the other possibility in the non-standard cosmology,
namely  large individual flavor lepton asymmetry
is presented when AMP, $e.g.$, the sneutrino  freeze-out (below $T_{sp}$ so  baryon asymmetry can be small).
As a consequence, the equilibrium with cosmic background does not imply
the wash-out effects at all.

\section*{Acknowledgement}

We would like to thank Chaoqiang Geng for helpful discussions.
 ZK thanks NCTS, Taiwan for hospitality during the early stage of this work.
This research was supported in part
by the Natural Science Foundation of China
under grant numbers 10821504, 11075194, 11135003, and 11275246,
and by the DOE grant DE-FG03-95-Er-40917 (TL).

\section*{Note added}

After the completion of this work, we noticed the paper~\cite{arXiv:1111.5615}, which also
studied the Majorana LSP from the AMP late decays in the supersymmtric models.
However, they considered the Bino dark matter.


\end{document}